\documentclass[conference]{IEEEtran}
\IEEEoverridecommandlockouts
\usepackage[utf8]{inputenc}
\usepackage{graphicx}
\usepackage{amsmath}
\usepackage{color}

\begin{document}
\title{ArrayFlex: A Systolic Array Architecture with Configurable Transparent Pipelining}

\author{
\IEEEauthorblockN{C. Peltekis, D. Filippas, G. Dimitrakopoulos}
\IEEEauthorblockA{\footnotesize Electrical and Computer Engineering\\ 
Democritus University of Thrace, Greece}
\and
\IEEEauthorblockN{C. Nicopoulos}
\IEEEauthorblockA{\footnotesize Electrical and Computer Engineering\\  University of Cyprus, Cyprus}
\and
\IEEEauthorblockN{D. Pnevmatikatos}
\IEEEauthorblockA{\footnotesize Electrical and Computer Engineering\\  
National Technical University of Athens, Greece}}
\maketitle

\maketitle

\begin{abstract}
Convolutional Neural Networks (CNNs) are the state-of-the-art solution for many deep learning applications. For maximum scalability, their computation should combine high performance and energy efficiency. In practice, the convolutions of each CNN layer are mapped to a matrix multiplication that includes all input features and kernels of each layer and is computed using a systolic array. 
In this work, we focus on the design of a systolic array with configurable pipeline with the goal to select an optimal pipeline configuration for each CNN layer.
The proposed systolic array, called ArrayFlex, can operate in normal, or in shallow pipeline mode, 
thus balancing the execution time in cycles and the operating clock frequency. By selecting the appropriate pipeline configuration per CNN layer, ArrayFlex reduces the inference latency of state-of-the-art CNNs by 11\%, on average, as compared to a traditional fixed-pipeline systolic array. Most importantly, this result is achieved while using 13\%--23\% less power, for the same applications, thus offering a combined energy-delay-product efficiency between 1.4$\times$ and 1.8$\times$.
\end{abstract}

\section{Introduction}
\label{s:intro}

The quality of deep learning has increased significantly with Convolutional Neural Networks (CNNs)~\cite{convnext}. CNNs have enabled remarkable performance in many application fields, such as computer vision~\cite{convnext, mobilenet}, natural language processing~\cite{NLP_CNN}, and robotics~\cite{CNN-SLAM}. 

This widespread adoption of CNNs has triggered the need to accelerate them directly in hardware. To do so, CNN layers are mapped to General Matrix Multiplication (GEMM) kernels~\cite{cudnn}. GEMMs are at the heart of deep learning hardware and they naturally map onto Systolic Arrays (SA)~\cite{why-systolic}. Tensor-processing units~\cite{tpu} and other related architectures~\cite{scalesim, auto-sa, meissa, factored-sa} are characteristic examples of newly designed SAs.

Systolic arrays have also been implemented as configurable architectures that support arbitrary bit-width arithmetic precision to enable sub-word parallelism~\cite{reconfig-bitwidth}. Furthermore, configurable SAs can support various dataflows~\cite{eyriss2,hetero-sa} that are more amenable to the different forms of convolutions found across different CNNs, and even across different layers within one model. Coarse-grained SA reconfiguration allows for the partitioning of the hardware resources into multiple CNNs that execute concurrently on the different parts of the SA~\cite{dataflow_mirroring, planaria, sara}. In the same context, SAs have been customized to handle sparse matrices~\cite{sparse-tpu}.

In this work, we focus on customizing the \textit{pipeline} structure of SAs with the goal being to reduce the execution latency of matrix multiplication. Throughput can always be increased by adding more processing elements and increasing the input and output bandwidth of the SA. On the contrary, optimizing the execution \textit{latency} requires architectural re-organization of the SA that should also adjust to the structure of each CNN layer. Reducing latency is important in applications executed at the edge~\cite{edge} and a necessity for applications that also require real-time responses~\cite{edge-2}. Moreover, for small batch sizes, reducing the latency can also reduce the time to the final result. This is critical in RNNs, which are harder to batch than CNNs~\cite{rnn-batch}.

The proposed SA architecture, named \emph{ArrayFlex}, can configure its pipeline structure between normal and various shallow pipeline depths. In shallow mode, two or more adjacent pipeline stages are joined by bypassing intermediate pipeline stage(s)~\cite{collapse, transparent}. This merging effectively reduces the number of cycles needed to complete matrix multiplication. On the other hand, the clock frequency is reduced to avoid timing violations due to the increased logic depth. This double-faceted tradeoff allows us to identify the best possible configuration per CNN layer that minimizes the total execution latency in absolute time. Overall, the contributions of this work can be summarized as follows:
\begin{itemize}
\item ArrayFlex introduces a configurable pipeline architecture for SAs that can adjust its pipeline depth to the size of the corresponding matrix multiplication while aiming to minimize the total execution latency.
\item When shallow pipeline mode is beneficial, power is equivalently reduced, since transparent registers remain clock-gated and the design as a whole operates at a lower clock frequency.
\item Extensive evaluations using state-of-the-art CNN applications demonstrate that the proposed architecture reduces the latency by 11\%, on average, while also consuming 13\%--23\% less power, as compared to SAs with a fixed pipeline organization. This amounts to a substantial improvement in overall energy efficiency.
\end{itemize}

The rest of the paper is organized as follows: Section~\ref{s:gemm} revisits the basics of computing matrix multiplication on SAs. Section~\ref{s:transparent} introduces the proposed ArrayFlex SA architecture with configurable pipeline depth. Experimental results are presented in Section~\ref{s:eval} and conclusions are drawn in Section~\ref{s:conclusions}.

\section{Matrix multiplication on Systolic Arrays}
\label{s:gemm}

\begin{figure}[t]
\centering
\begin{tabular}{cc}
\includegraphics[width=0.44\columnwidth]{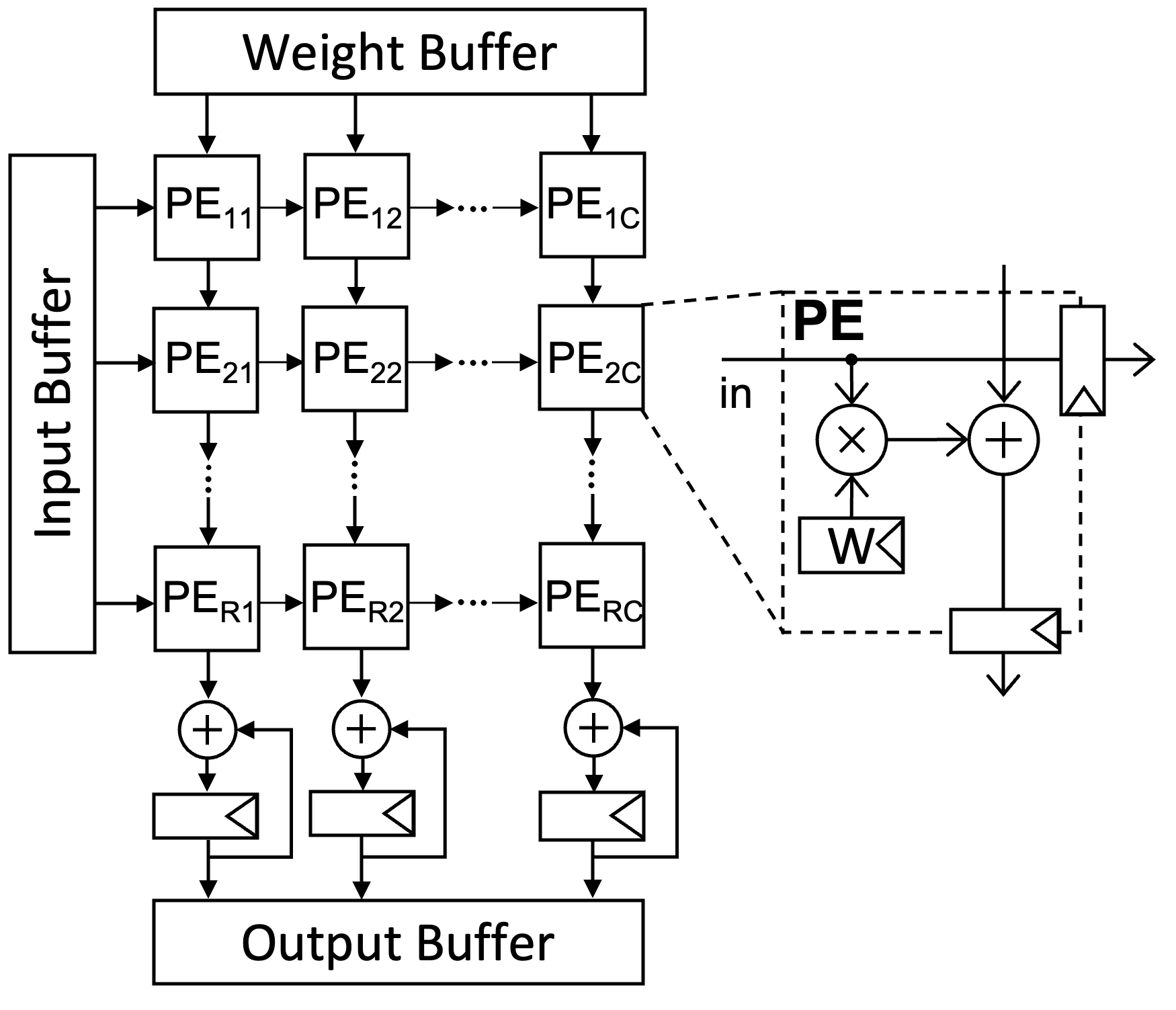} &
\includegraphics[width=0.40\columnwidth]{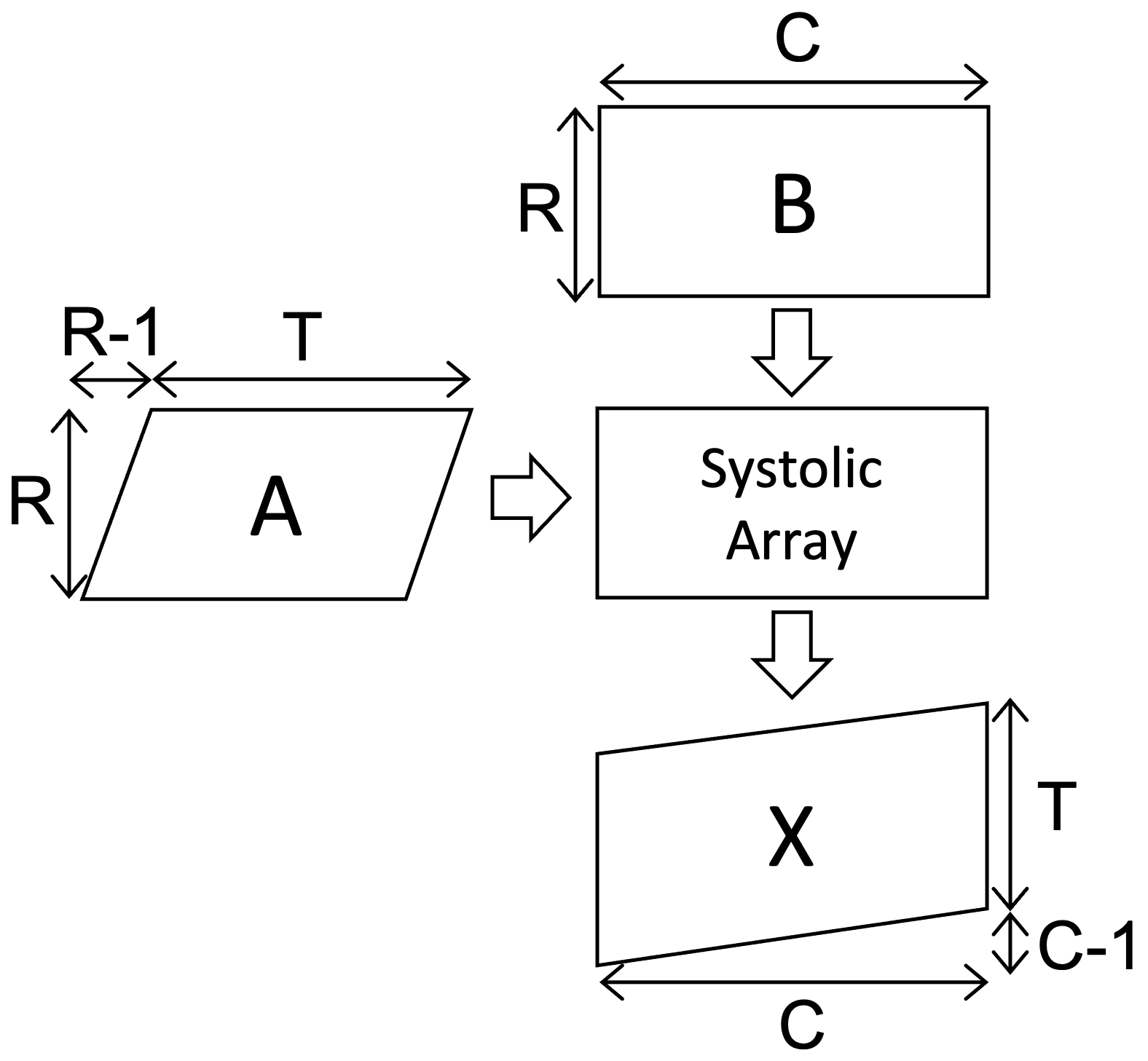} \\
{\small (a) Typical SA organization} & 
{\small (b) Weight-Stationary Dataflow} \\
\multicolumn{2}{c}{\includegraphics[width=0.5\columnwidth]{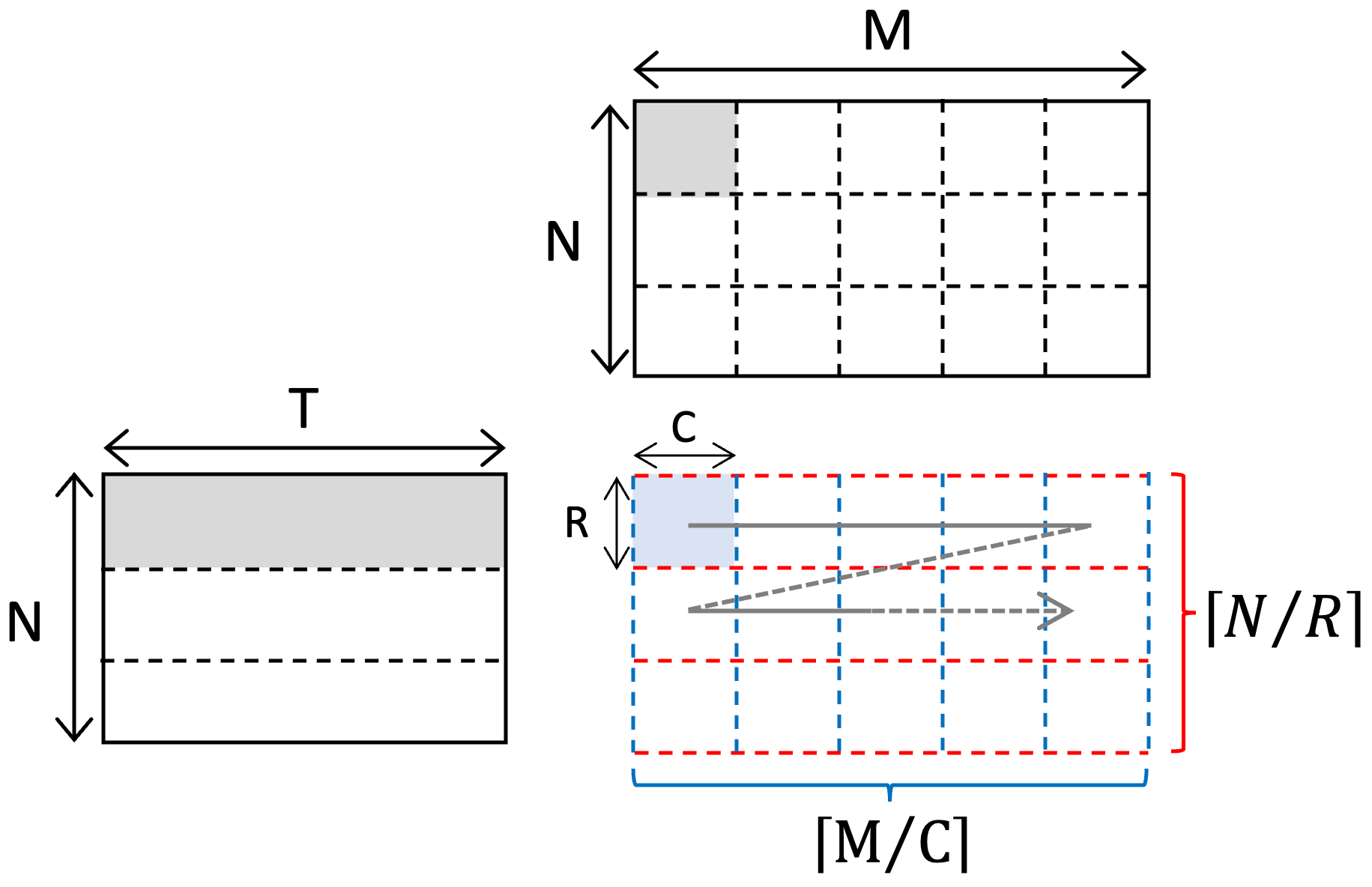}} \\
\multicolumn{2}{c}{{\small (c) Tiled matrix multiplication}}
\end{tabular}
\caption{The organization of a typical SA and the structures of the Weight-Stationary dataflow and of tiled matrix multiplication.} 
\label{f:ws-base}
\end{figure}

SAs consist of an array of Processing Elements (PEs) organized in $R$ rows and $C$ columns, as shown in Fig.~\ref{f:ws-base}(a). Each PE consists of a multiplier and an adder and necessary registers to appropriately pipeline the operation. SAs are fed by local memory banks placed on the west edge (for the input features) and the north edge (for the weights of the convolution kernels) of the array, while output results are collected on the south edge. 
The dataflow selected for the SA determines the structure of the PEs and how matrix multiplication $A\times B$ is executed.
Fig.~\ref{f:ws-base}(b) shows how the SA performs matrix multiplication using the Weight-Stationary (WS) dataflow~\cite{scalesim}. WS is generally preferred over other dataflows, since it exploits high spatio-temporal reuse of the weights~\cite{tpu, meissa}.

Matrix multiplication $X_{T, M}\! =\! A_{T, N} \!\times\! B_{N, M}$ can be mapped to a SA, if the SA is large enough to accommodate in parallel a column of $A$ and a row of $B$. This can happen when the spatial dimensions $N$ and $M$ of matrix multiplication match the size of the SA, i.e., $R = N$ and $C = M$. 

In this case, and assuming a WS dataflow, as shown in Fig.~\ref{f:ws-base}(b), matrix $B$ is first pre-loaded in the SA by loading a new row per cycle. Thus, $R$ cycles are needed to complete the loading. 
Once $B$ is loaded, matrix $A$ is streamed in from the left edge of the array. The first arriving element would reach the rightmost column of the SA after $C-1$ cycles. After the top row is filled with the incoming data elements, it takes $R-1$ cycles to reduce the result of all PEs of the same column. For the reduction operation, the result of the multiplication and addition in each PE is first registered at the borders of the PE and it then moves downwards to the next PE of the same column.
The SA becomes empty when the reduction is finished on the rightmost column for all incoming skewed columns of $A$. Since $A$ consists of $T$ rows, the overall latency $L$ of computing matrix multiplication equals:
\begin{equation}
L = 2R + C + T - 2
\label{e:latency-base}
\end{equation}

When the size of matrices $A$ and $B$ is larger than the size of the SA, i.e., $N > R$ and/or $M > C$, 
matrix multiplication is executed in tiles, as shown in Fig.~\ref{f:ws-base}(c), where each tile (sub-matrix) matches the size of the SA. 
The partial sums for each tile that reach the bottom of the SA, get accumulated into the corresponding output accumulator located below the SA (see Fig.~\ref{f:ws-base}(a)). 
According to~\cite{scalesim}, the latency of
tiled matrix multiplication is equal to 
\begin{equation}
L_{\text{total}} = L \times \left\lceil \frac{N}{R} \right\rceil \times \left\lceil \frac{M}{C} \right\rceil.
\label{e:latency-total-base}
\end{equation}
$L$ is the latency for computing the product of two tiles
$A^{\text{sub}}_{T, R}\times B^{\text{sub}}_{R, C}$ 
and it is given by~\eqref{e:latency-base}, and $\lceil \frac{N}{R} \rceil \times \lceil \frac{M}{C} \rceil$ represents the total number of tiles.

\section{A Systolic Array with Configurable Transparent Pipelining}
\label{s:transparent}

In this work, we aim to adjust the pipeline depth of the vertical and horizontal pipelines of the SA, in order to optimally calibrate them to the size of the systolic array ($R$ and $C$), and the size $T$ of matrix $A$, which are crucial in the overall latency of computation.

\begin{figure}[thb]
\centering
\includegraphics[width=0.78\columnwidth]{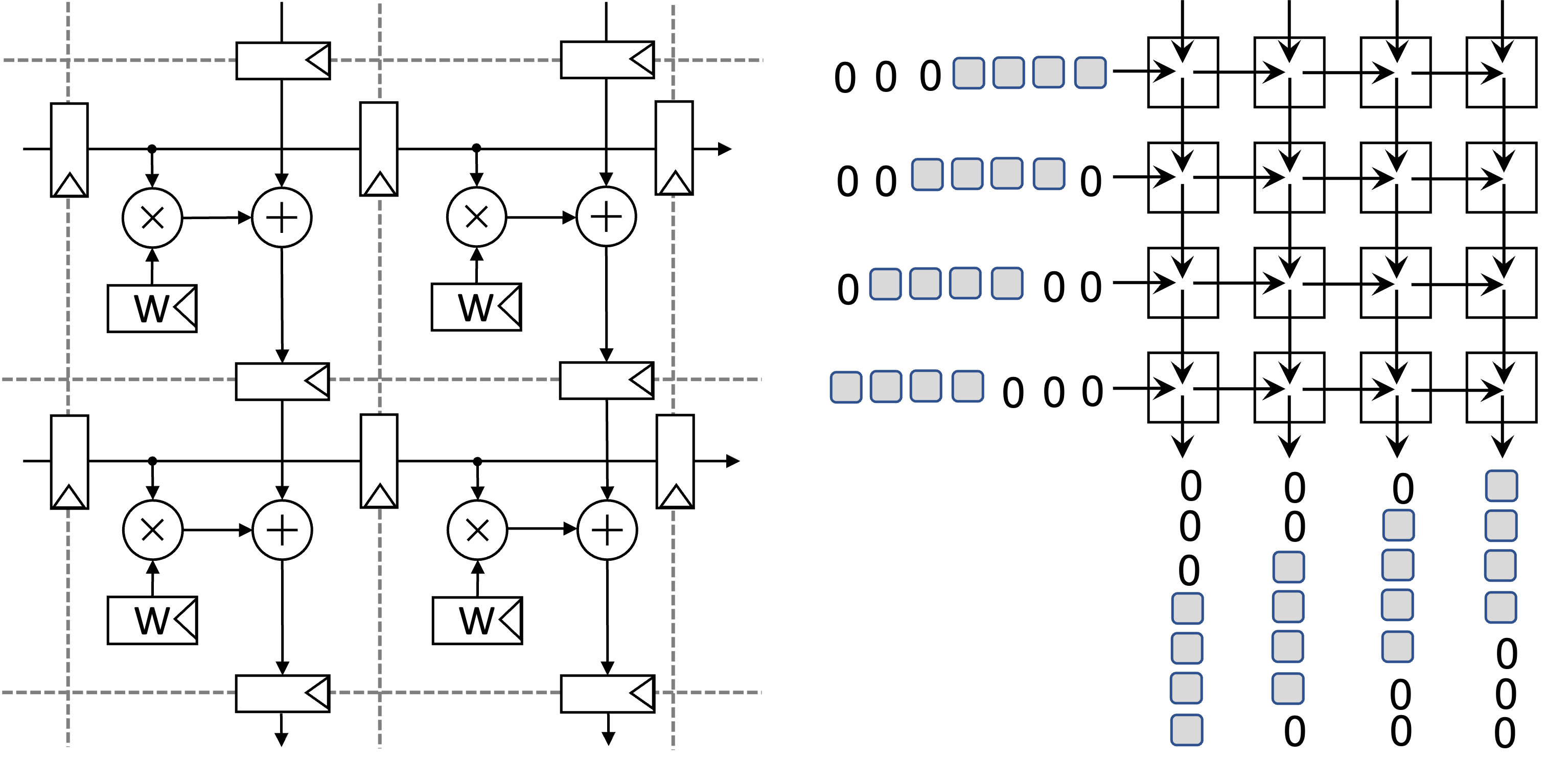} \\
{\small (a) Normal pipeline; $k=1$} \\
\vskip 0.2cm
\includegraphics[width=0.78\columnwidth]{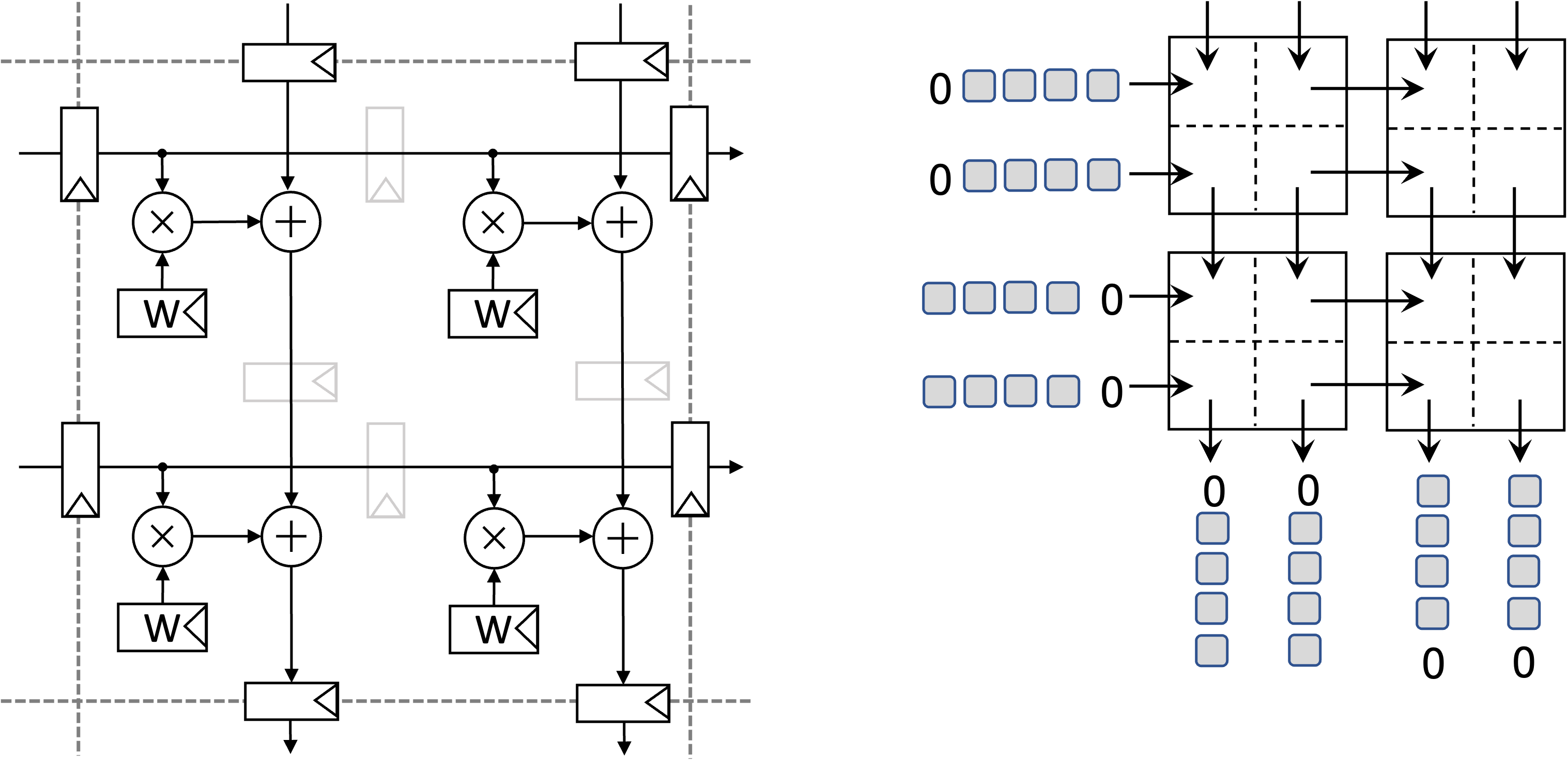} \\
{\small (b): Shallow pipeline; $k=2$} 
\caption{(a) In normal pipeline mode, each data item moves to the next PE, either horizontally or vertically, in one clock cycle. (b) In shallow pipeline mode,
the dataflow of every $k=2$ PEs is merged in a single-cycle operation. Merging is possible by bypassing the intermediate pipeline registers. The input and output dataflow skew is altered to match the shallower pipeline structure.}
\label{f:transparent}
\end{figure}

To reduce the $R-1$ cycles spent in the reduction operation in each column of the SA, we can configure the vertical reduction pipeline to operate in shallow mode, where two or more adjacent pipeline stages are merged by making the intermediate registers transparent. The registers in transparent mode, bypass the input data to the next stage, thereby joining two adjacent combinational logic circuits into one pipeline stage. Up to $k$ registers can be joined in the vertical direction.

Pipeline collapsing is also performed in the horizontal dataflow.
Instead of letting the input stream move one column to the right in each cycle, we allow it to broadcast to $k$ columns when operating in shallow pipeline mode~\cite{risset}.

The normal pipeline mode that corresponds to the case of $k=1$ is shown in Fig.~\ref{f:transparent}(a). Similarly, Fig.~\ref{f:transparent}(b) depicts an example of shallow pipeline operation assuming $k=2$. In this case, the result of the top row of PEs is added transparently to the result of the second row of PEs in the same clock cycle. The same operation occurs for every two adjacent PEs.

To align the shallow pipelines of the SA with the arrival of the input data, their arrival skew should be altered. The first (and last) elements of matrix $A$ arrive in batches of $k$ words. It should be stressed that this change does not fundamentally alter the operation of the systolic array, since the required input and output bandwidth remains the same and it is equal to $R$ and $C$ words per cycle (i.e., equal to the number of rows and columns of the SA).

\subsection{Latency vs. clock frequency tradeoff}

Using this approach, and assuming that we can collapse/merge $k$ intermediate PEs into the same pipeline stage in both the vertical and the horizontal directions, the number of cycles spent in the reduction operation reduces from $R-1$ to $\frac{R}{k} - 1$, and the number of cycles spent in the broadcast of the first data element to the rightmost column of the SA reduces from $C-1$ to $\frac{C}{k}-1$. Thus, the overall latency of computing a matrix product $A^{\text{sub}}_{T, R}\times B^{\text{sub}}_{R, C}$ (as needed by each sub-matrix of the original $A\times B$ product) becomes
\begin{equation}
L(k) = R + \frac{R}{k}+\frac{C}{k}+T-2 
\label{e:lat-new}
\end{equation}
The number of cycles needed for all tiles is then equal to 
\begin{equation}
L_{\text{total}}(k) = L(k) \times \left\lceil \frac{N}{R} \right\rceil \times \left\lceil \frac{M}{C} \right\rceil.
\label{e:lat-total-new}
\end{equation}
Overall, the higher the amount of collapsing (i.e., higher value of $k$), the larger the reduction in the number of cycles needed to complete matrix multiplication.

On the other hand, to enable pipeline collapsing within the SA, one must slow down the clock frequency to avoid timing violations due to increased logic depth. Column collapsing only affects the delay marginally. However, row collapsing requires $k$ additions to be performed in series in the same clock cycle. 
Therefore, for each shallow pipeline configuration, there is an equivalent throttling of the operating clock frequency to adjust the clock period to the logic depth of the selected configuration. Our goal is to select the best possible $k$ that minimizes the total execution latency in absolute time (i.e., clock cycles $\times$ clock period), given the size of the systolic array $R\times C$ and the size of the matrix multiplication, as determined by $N, M$, and $T$.

\subsection{The Organization of Configurable PEs}

The minimum clock period that the design can operate at is determined by the maximum logic delay between any two pipeline registers, plus any clocking overhead (sum of the register clock-to-Q delay and the setup time of the flip-flops). In the baseline case ($k=1$), the maximum combinational delay remains inside the borders of one PE and is equal to the delay of the multiplier and the delay of the adder.

\begin{figure}
\centering
\includegraphics[width=0.45\columnwidth]{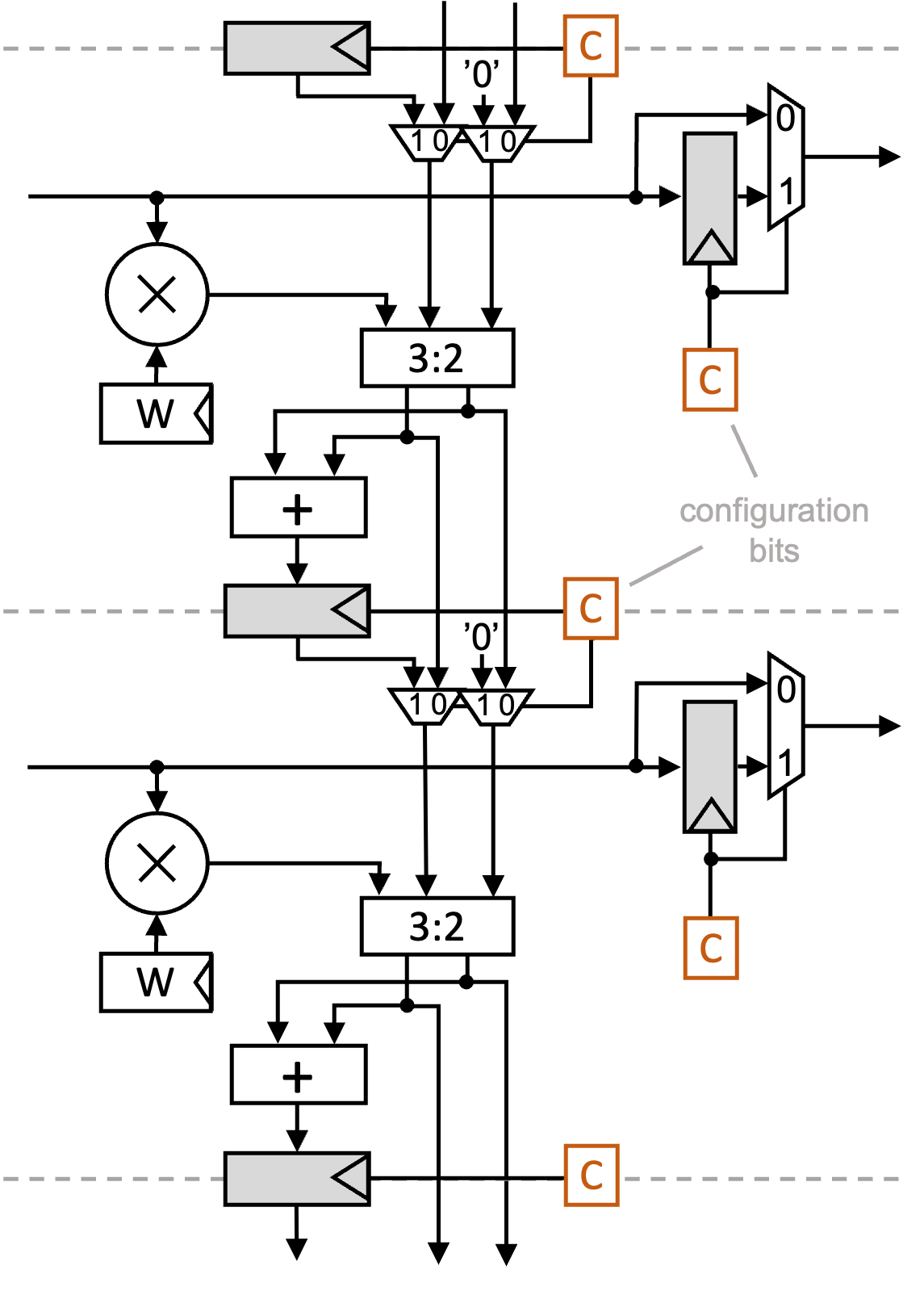}
\caption{The organization of two enhanced, configurable PEs of the same column. Registers are bypassed in the vertical and horizontal directions according to the pipeline configuration. In shallow pipeline mode, reduction is performed using a series of 3:2 carry-save adders ending with a carry-propagate adder.}
\label{f:new-pe}
\end{figure}

When collapsing $k$ pipeline stages into one, the maximum combinational delay again involves the delay of one multiplier plus the delay of $k$ carry-propagate adders in series plus the delay of bypass multiplexers. To avoid this significant delay overhead when collapsing adjacent pipeline stages, we augment the PEs of the SA with an additional 3:2 carry-save stage that is only enabled during shallow pipeline mode. The organization of two enhanced PEs of the same column are shown in Fig.~\ref{f:new-pe}. The 3:2 carry-save adder is composed of parallel full-adders, with each one placed at each bit position.

When PEs are collapsed, the registers placed in the horizontal direction are bypassed (and clock gated) by additional multiplexers controlled by configuration bits loaded in parallel to the weights of matrix $B$. In the vertical direction, to add the $k$ products produced by the multipliers of each PE, we utilize $k$ carry-save adder stages. The carry-save adders are connected in series through additional bypass multiplexers placed in the vertical direction. At the last stage, where pipeline collapsing ends and the result needs to be saved in the corresponding pipeline register, the result of the carry-save adders is transformed to one operand using the carry-propagate adder of each PE.

\begin{figure}[htb]
\begin{tabular}{cc}
\includegraphics[width=0.31\columnwidth]{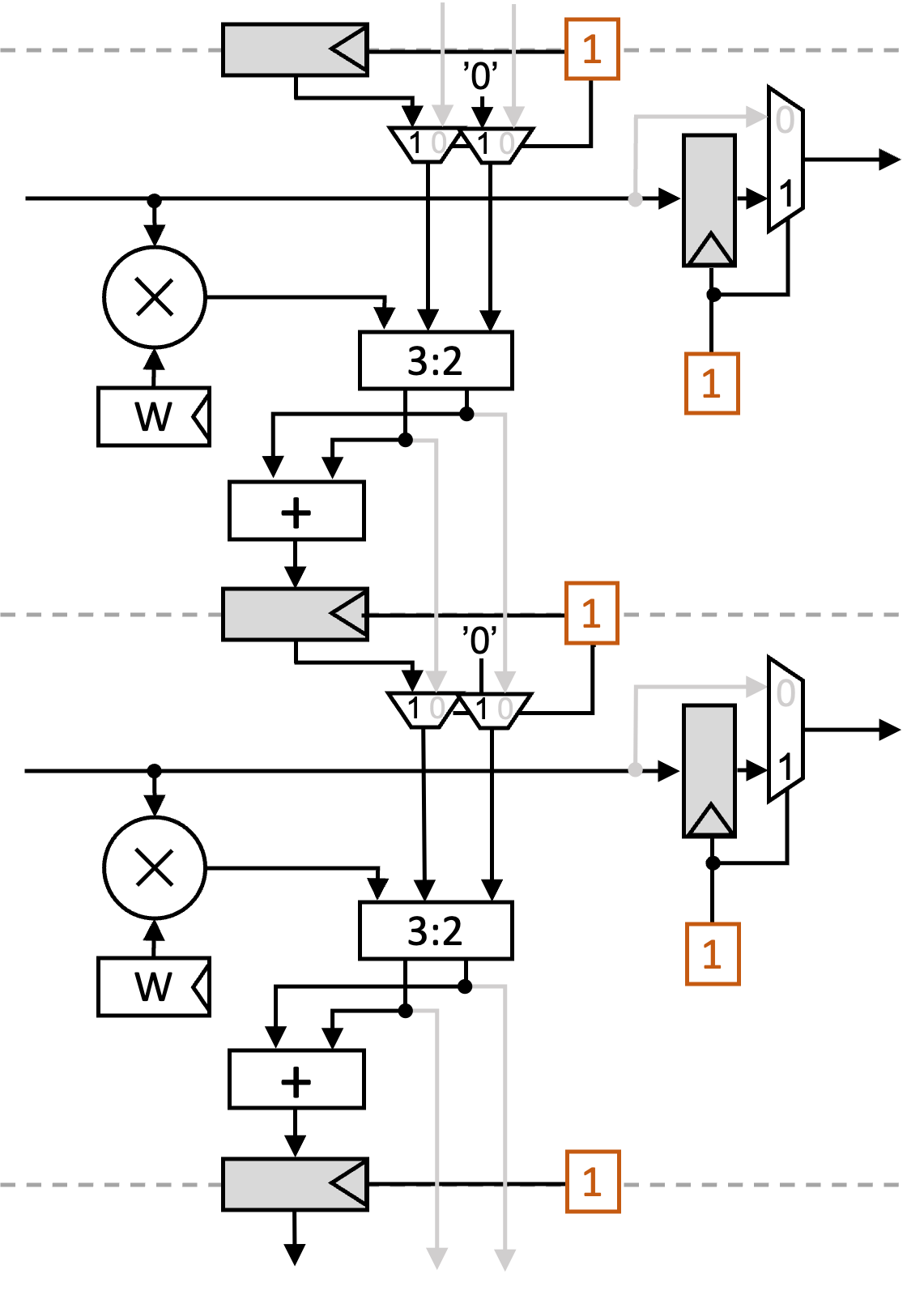} 
 &
\includegraphics[width=0.58\columnwidth]{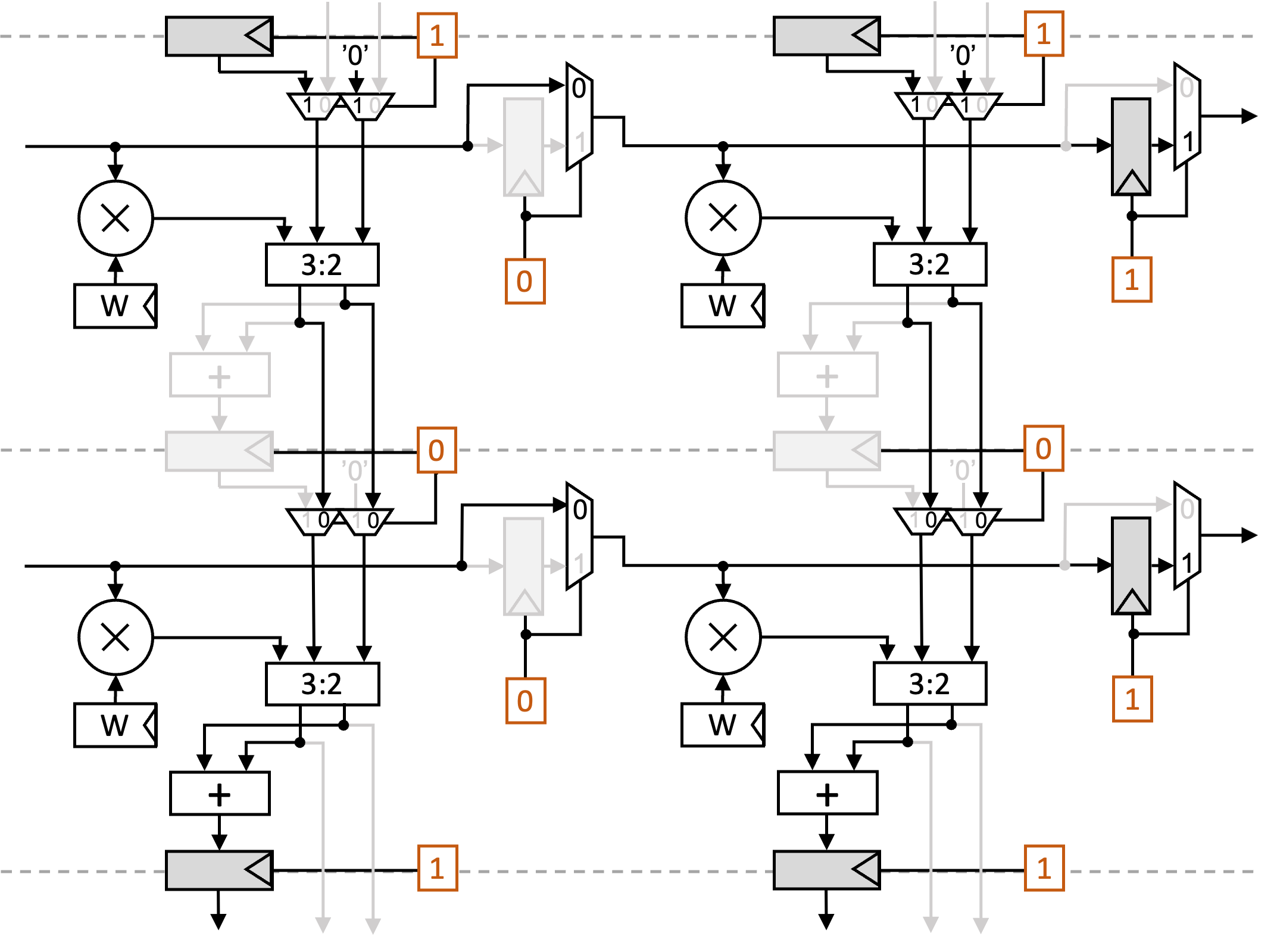}
\\
(a) Normal pipeline & (b) Shallow pipeline; $k=2$ \\ 
\end{tabular}
\caption{Example of active paths for (a) a normal pipeline ($k=1$), and (b) a shallow pipeline ($k=2$).}
\label{f:transparent-example}
\end{figure}

This configuration is shown in Fig.~\ref{f:transparent-example}(b) for $k=2$. The products of the first and the second row in each column are added using carry-save adders. The final sum is produced by the carry propagate adders of the last row. The carry propagate adders of the top row are not used, while the registers that are bypassed are clock-gated to save power. 

Each PE needs two configuration bits that configure independently the transparency (bypassing) of the pipeline registers in each direction.
Separate configuration bits per PE are needed since each PE can play a different role depending on the selected pipeline mode.

The incoming input and the weight stored in each PE have the same bit width. However, the vertical connections, including the carry-save adders and the carry propagate adders, have double the bit width, in order to accommodate the full product of the multiplier.

The 3:2 carry-save adder and the bypass multiplexers participate in the operation of each PE even when configured for a normal pipeline, i.e., $k=1$. As shown in Fig.~\ref{f:transparent-example}(a), the product of each multiplier is first added to the result of the previous PE of the same column through the carry-save adder before finalizing the addition with the carry-propagate adder. This extra hardware placed in series between the multiplier and the adder inevitably affects the minimum delay that can be achieved by a conventional PE that does not offer any pipeline reconfiguration. The experimental results show that this delay overhead is marginal and does not limit the applicability of the proposed approach.

\subsection{Minimizing the total execution time}

To identify the optimal value of $k$ that best fits the examined configuration, we first need to develop for ArrayFlex a rough model of how the clock period is affected with respect to $k$.

When collapsing $k$ pipeline stages of the SA, the maximum combinational delay involves the delay of $k$ bypass multiplexers in the horizontal direction, the delay of the multiplier ($d_{\text{mul}}$) of the rightmost PE of the collapsed pipeline block, plus the delay of $k$ cascaded 3:2 carry-save adders ($d_{\text{CSA}}$) and bypass multiplexers ($d_{\text{mux}}$) in the vertical direction. In this delay, we should add the delay of the final carry propagate adder ($d_{\text{add}}$) of the last row of the collapsed pipeline block and any flip-flop clocking overhead ($d_{\text{FF}}$). Overall, we can roughly estimate that the minimum clock period that can be achieved by a $k$-collapsed pipeline is:
\begin{equation}
T_{clock}(k) = d_{\text{FF}} + d_{\text{mul}} + d_{\text{add}} + 
k ( d_{\text{CSA}} + 2d_{\text{mux}} )
\label{e:clock}
\end{equation}

In practice, the design supports a maximum pipeline collapsing depth $k_{\text{max}}$. When collapsing fewer than $k_{\text{max}}$ pipeline stages, the combinational paths that still exist in the design but are not used are considered false paths. We provide this information explicitly to the static timing analyzer.

The latency in absolute time $T_{abs}(k)$ of computing a complete matrix multiplication using an SA with $k$-collapsible pipeline is the product of the latency in clock cycles $L_{\text{total}}(k)$ given in Equation~\eqref{e:lat-total-new} and the minimum clock period $T_{\text{clock}}(k)$ that corresponds to each $k$ (as given by Equation~\eqref{e:clock}): 
\begin{equation}
T_{\text{abs}}(k) = L_{\text{total}}(k)\times T_{\text{clock}}(k)
\end{equation}

\begin{figure}[t]
\centering
\begin{tabular}{cc}
\includegraphics[width=0.35\columnwidth]{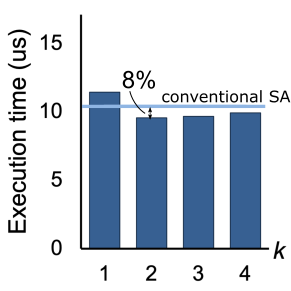} &
\includegraphics[width=0.35\columnwidth]{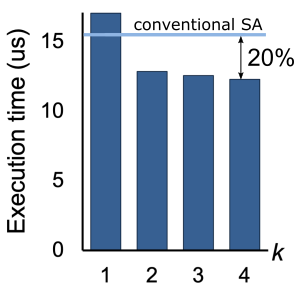} \\
{\small (a) ResNet-34 Layer 20} & 
{\small (b) ResNet-34 Layer 28} \\ 
\end{tabular}
\caption{The execution time of computing layers (a) 20 and (b) 28 of ResNet-34~\cite{resnet} as an equivalent matrix multiplication using a configurable SA under various pipeline collapsing depths $k$. The execution time of the conventional (non-configurable) SA that operates only in normal pipeline mode under the highest clock frequency is depicted as a straight line.}
\label{f:example}
\end{figure}

To explore the interesting interplay between computation latency in cycles and the configurable pipeline depth of the columns of the systolic array, we performed a simple experiment. We measured the execution time required to compute layers 20 and 28 of ResNet-34 CNN~\cite{resnet} as matrix multiplications using a configurable SA that consists of $(R, C) = (132, 132)$ rows and columns. The values of $R$ and $C$ were selected to be divisible by all the examined values of $k$, i.e., 1, 2, 3, and 4. The sizes of the corresponding matrix multiplications for computing layers 20 and 28 are $(M, N, T)=(256, 2304, 196)$ and $(M, N, T)=(512, 2304, 49)$, respectively. In both cases, we examined various pipeline collapsing depths. The obtained results in each case are depicted in Fig.~\ref{f:example}.

For each pipeline collapsing depth $k$, we scaled the clock frequency accordingly to match the combinational delay for each case. The execution latencies that correspond to a conventional (non-configurable) SA are shown as straight lines in both cases of Fig.~\ref{f:example}. The conventional SA operates using a normal pipeline at the highest clock frequency, since it does not suffer any delay overhead associated with configurability.

According to Fig.~\ref{f:example}(a), the execution time for layer 20 is minimized at $k=2$. In this case, the reduction of clock cycles and the increase of the clock period find their optimal match. Collapsing the pipeline deeper, i.e., $k=3$, still reduces the execution time, relative to a conventional SA, but the savings are less. For layer 28, as depicted in Fig.~\ref{f:example}(b), deeper pipeline collapsing offers the best execution time. In this case, utilizing a pipeline collapse depth of $k=4$ is the best choice.

To identify the optimal pipeline depth $\hat{k}$ that minimizes $T_{\text{abs}}(k)$, 
we take the derivative of $T_{\text{abs}}(k)$ with respect to $k$
and set it equal to zero. This leads to:
\begin{equation}
\hat{k} = \sqrt{\left ( \frac{R+C}{R+T-2}\right) \left(\frac{d_{\text{FF}} + d_{\text{mul}} + d_{\text{add}}}{d_{\text{CSA}} + 2d_{\text{mux}}}\right )}
\label{e:opt-k}
\end{equation}
Even though $k$ is a discrete variable, Eq.~\eqref{e:opt-k} gives a simple analytical model that leads to one interesting conclusion: the pipeline depth of the SA should be judged not only on the delay profile of the adder and the multiplier hardware blocks, but also on the size of the matrix multiplication (dimension $T$) relative to the size of the SA.

For instance, the first layers of the CNN try to identify on the input coarse features using a wide search area. This leads to large values for dimension $T$ on the corresponding matrix multiplication. As a result, $\hat{k}$ cannot easily reach values larger than one. This means that the best choice is to use an architecture with a normal pipeline, i.e., with $k=1$. On the contrary, in the last CNN layers, it is common the size of the input features to decrease and their number to increase~\cite{convnext, mobilenet}.
Effectively, in these layers the value of $T$ drops and using shallow pipelining (higher $k$) is a better choice. Allowing for pipeline collapse, which also reduces the clock frequency, not only reduces the overall execution time, but it also saves dynamic power. Under shallow pipelining, the clocking power is also reduced, since more registers remain clock-gated.

\section{Evaluation}
\label{s:eval}

In the experimental evaluation, the goal is to highlight: (a) when SAs operating in shallow mode make sense, (b) the latency and power savings in these cases, and (c) the area overhead incurred in offering pipeline-depth reconfigurability.

To answer these questions, we developed parameterized models of a conventional SA and ArrayFlex in SystemVerilog RTL. Both SAs operate on 32-bit quantized inputs and weights executing single-batch inference of various CNNs that consist of matrix multiplications of different sizes. The additions in each column of the SAs are performed at 64 bits.

The SAs were implemented using Cadence's digital implementation flow using a 28 nm standard-cell library. Conventional SAs operating only with a normal pipeline in a non-configurable manner can reach a clock frequency of 2 GHz. The proposed configurable SAs support one normal and two shallow pipeline modes. In normal pipeline mode ($k=1$), the proposed SA operates at 1.8 GHz. The two shallow pipeline modes allow for collapsing $k=2$ or $k=4$ pipeline stages. In these cases, the clock frequency is configured at 1.7 GHz and 1.4 GHz, respectively. Collapsing three pipeline stages is not supported, since three does not divide exactly with the size of the SA, which is a power-of-two in both dimensions.

\begin{figure}[ht]
\centering
\begin{tabular}{cc}
\includegraphics[width=0.32\columnwidth]{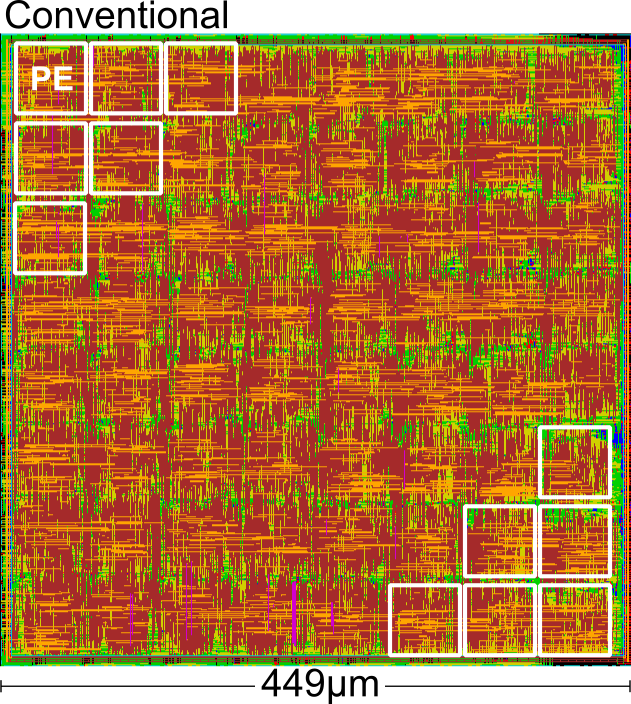} &
\includegraphics[width=0.3424\columnwidth]{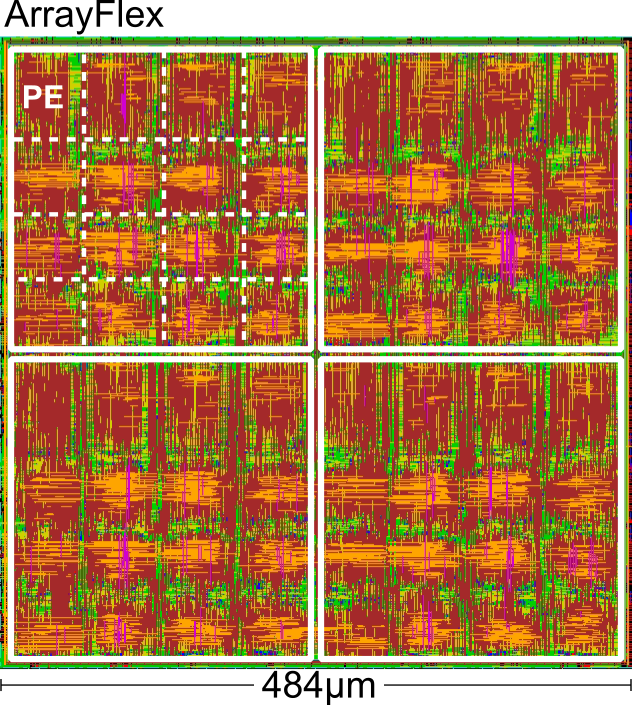} \\
\end{tabular}
\caption{The physical layouts of 8$\times$8 PEs using the conventional SA (left) and the proposed ArrayFlex design (right).}
\label{f:layout}
\end{figure}

To estimate the area cost of reconfigurability, Fig.~\ref{f:layout} highlights the physical layout of a conventional SA, relative to the ArrayFlex design, using 8$\times$8 PEs. From the physical layout of both SAs, it is evident that the area of ArrayFlex is increased in both dimensions. The area overhead per PE for this design is approximately 16\%. This extra area is consumed by the carry-save adder and the bypass multiplexers, while some marginal area is consumed by the two configuration bits per PE.

\begin{figure*}
\centering
\includegraphics[width=0.94\textwidth]{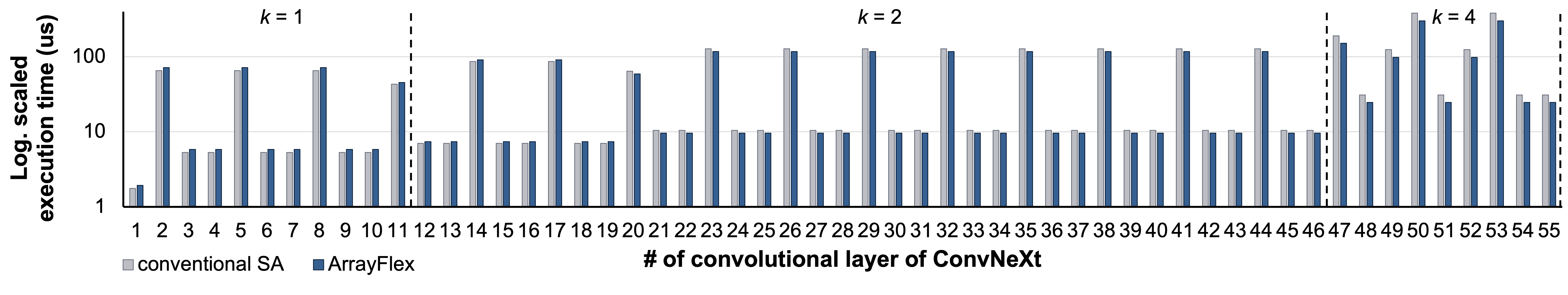}
\caption{The execution time of each CNN layer of ConvNeXt~\cite{convnext} using the conventional and the proposed ArrayFlex SAs. Size of both SAs: 128$\times$128 PEs.}
\label{f:convnext}
\end{figure*}

\subsection{Performance evaluation}

Initially, the aim is to highlight the effectiveness of configuring the pipeline depth per CNN layer in a way that minimizes the total execution time. Fig.~\ref{f:convnext} illustrates the execution time per CNN layer of ConvNeXt~\cite{convnext} using SAs that consist of 128$\times$128 PEs. The proposed ArrayFlex SA selects the optimal pipeline depth based on the structure of each CNN layer. For the first 11 layers, it is advantageous to operate under normal pipeline mode. This means that both the conventional SA and ArrayFlex require the same number of cycles to finish the matrix multiplication of each layer. Thus, since the conventional SA operates at a higher clock frequency, it finishes earlier in these cases. For layers 12--46, the proposed SA works optimally under a shallow pipeline mode of $k=2$, while, for layers 47--55, $k=4$ is the best configuration. In those cases, the execution time required by ArrayFlex is less than the execution time on the conventional SA. Interestingly, the best pipeline organization per CNN layer is approximated fairly accurately (assuming continuous values) by Equation~\eqref{e:opt-k}. 
For ArrayFlex, the execution time savings per layer range between 1.5\% and 26\%, while the \textit{total execution time} \textit{for all layers} is 11\% less than the time required by the conventional SA.

\begin{figure}[htb]
\centering
\begin{tabular}{cc}
\includegraphics[width=0.44\columnwidth]{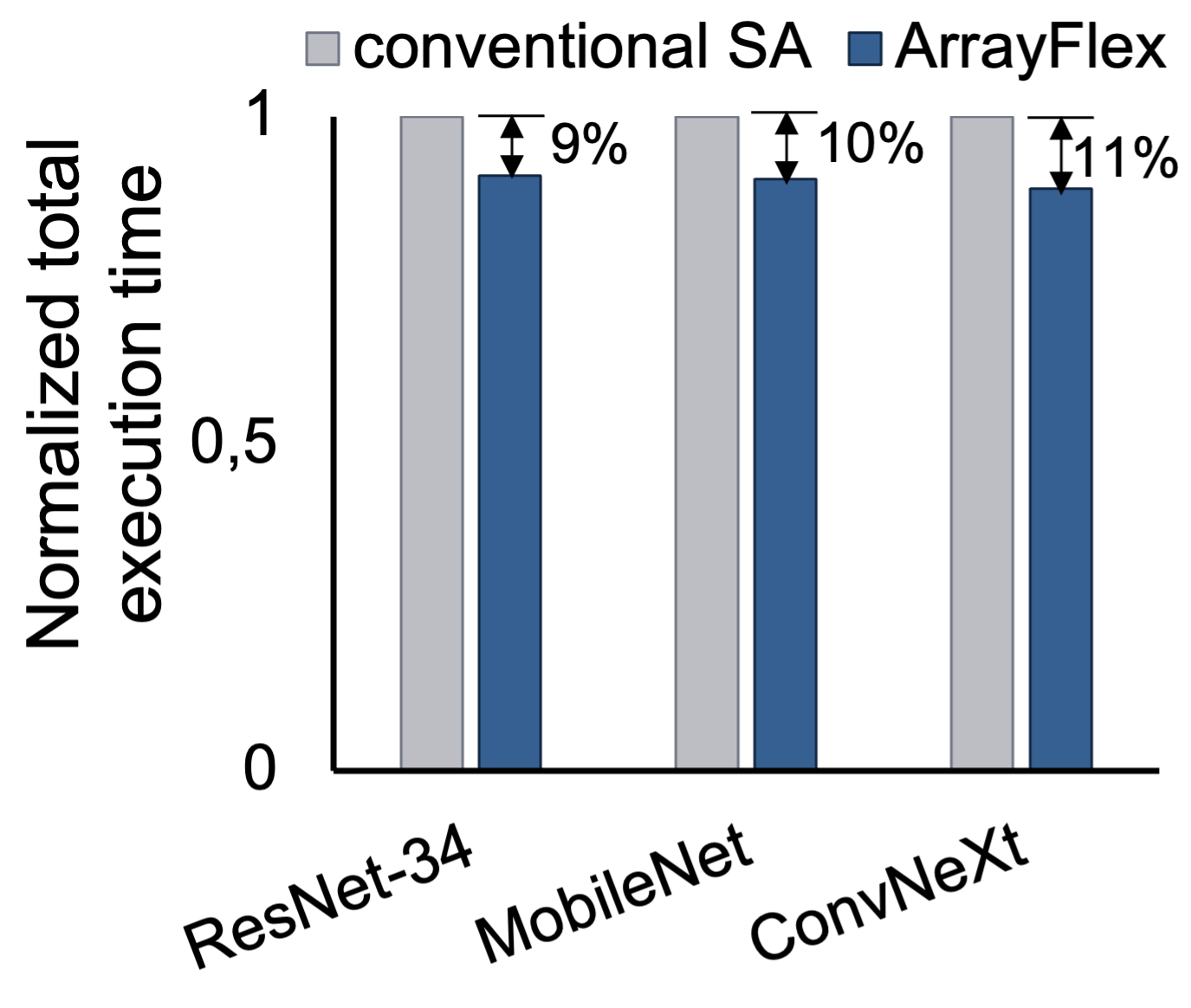} &
\includegraphics[width=0.44\columnwidth]{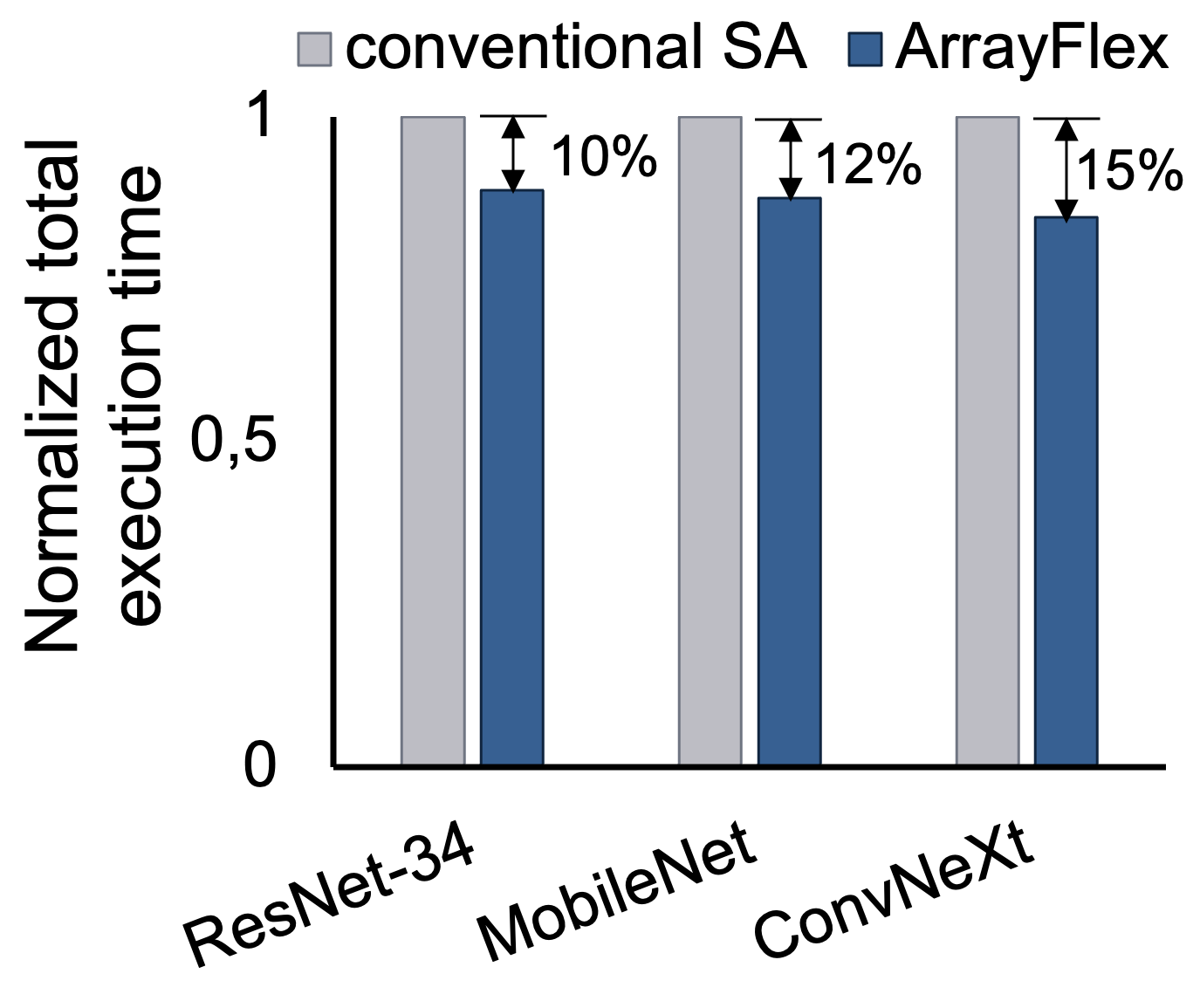} \\
{\small (a) 128$\times$128 SAs} &
{\small (b) 256$\times$256 SAs} 
\end{tabular}
\caption{The normalized execution times for \textit{complete runs} (i.e., execution of all layers) for three CNNs using (a) 128$\times$128 and (b) 256$\times$256 SAs. The times are normalized for visual clarity, since the execution time of ConvNeXt is significantly higher than the execution times of the other two CNNs.}
\label{f:apps}
\end{figure}

Similar behavior is observed under other CNN models and different SA sizes. Fig.~\ref{f:apps} depicts the normalized total execution time of three CNNs, ResNet-34~\cite{resnet}, MobileNet~\cite{mobilenet}, and ConvNeXt~\cite{convnext}, using 128$\times$128 and 256$\times$256 SAs. In all cases, the proposed ArrayFlex design, which configures the pipeline depth and the corresponding clock frequency to the characteristics of each CNN layer, achieves lower execution latency, ranging between 9\% and 11\%. The savings increase for larger SAs, since more CNN layers prefer a shallow pipeline configuration with $k=4$. This behavior is in line with Equation~\eqref{e:opt-k} that "predicts" higher values for $\hat{k}$ when the size of the SA increases, i.e., with larger values of $R$ and $C$.

\subsection{Power consumption evaluation}

One other equally important attribute of the proposed ArrayFlex architecture is that it reduces execution time \textit{without} increasing power. 

ArrayFlex has larger switched capacitance than a conventional SA, due to the extra hardware required to enable pipeline-depth configurability. Furthermore, it operates at a lower clock frequency than a conventional SA in all pipeline modes. The latter property partially amortizes the power cost of the additional hardware. However, in normal pipeline mode, ArrayFlex still consumes more power than a conventional SA. This behavior changes when in shallow pipeline mode, whereby the clock frequency is further reduced and additional power is saved by the clock gating of the bypassed registers. Therefore, the power profile of ArrayFlex strongly depends on the selected pipeline mode, which is decided independently for each CNN layer.

\begin{figure}[htb]
\centering
\begin{tabular}{cc}
\includegraphics[width=0.44\columnwidth]{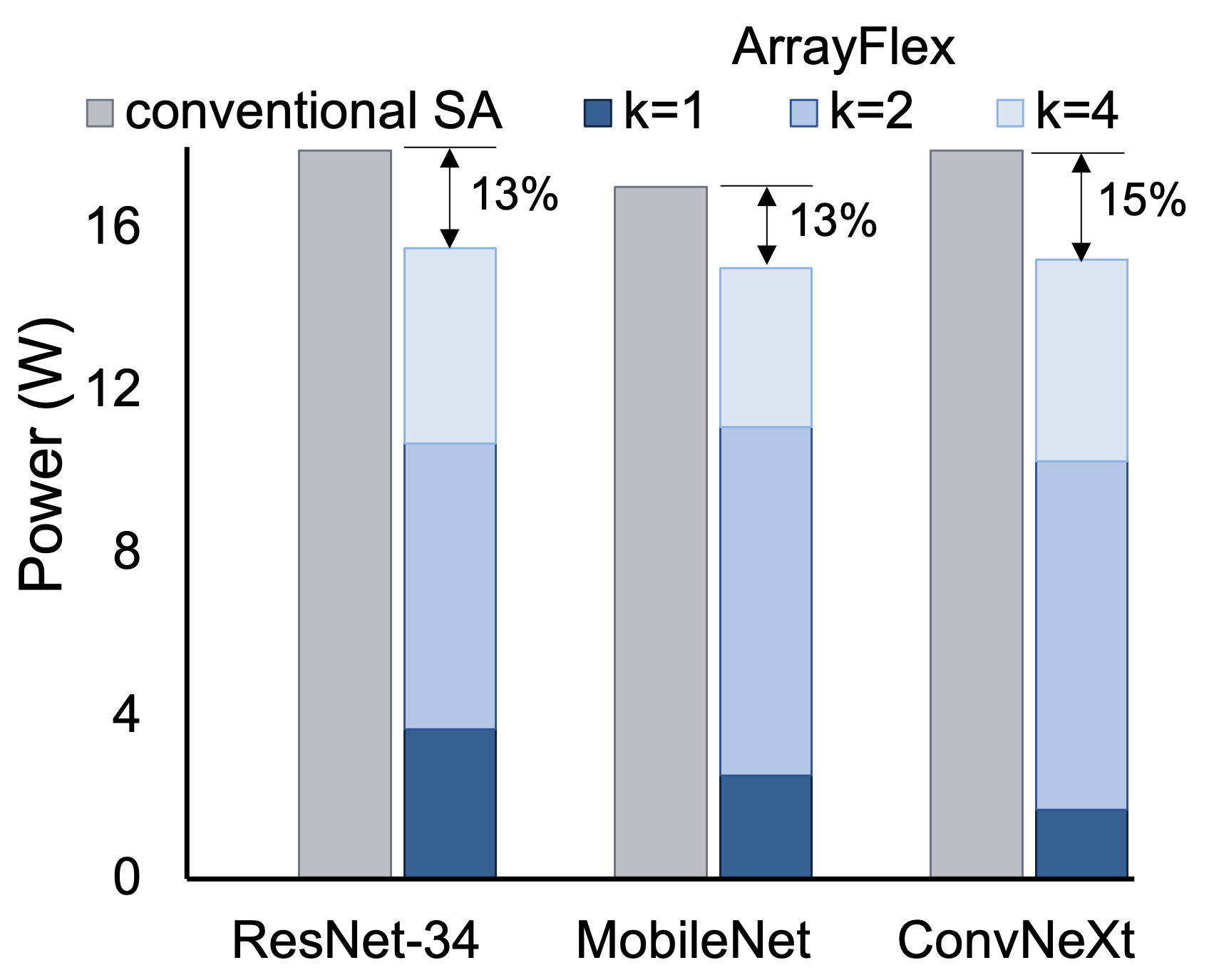} &
\includegraphics[width=0.44\columnwidth]{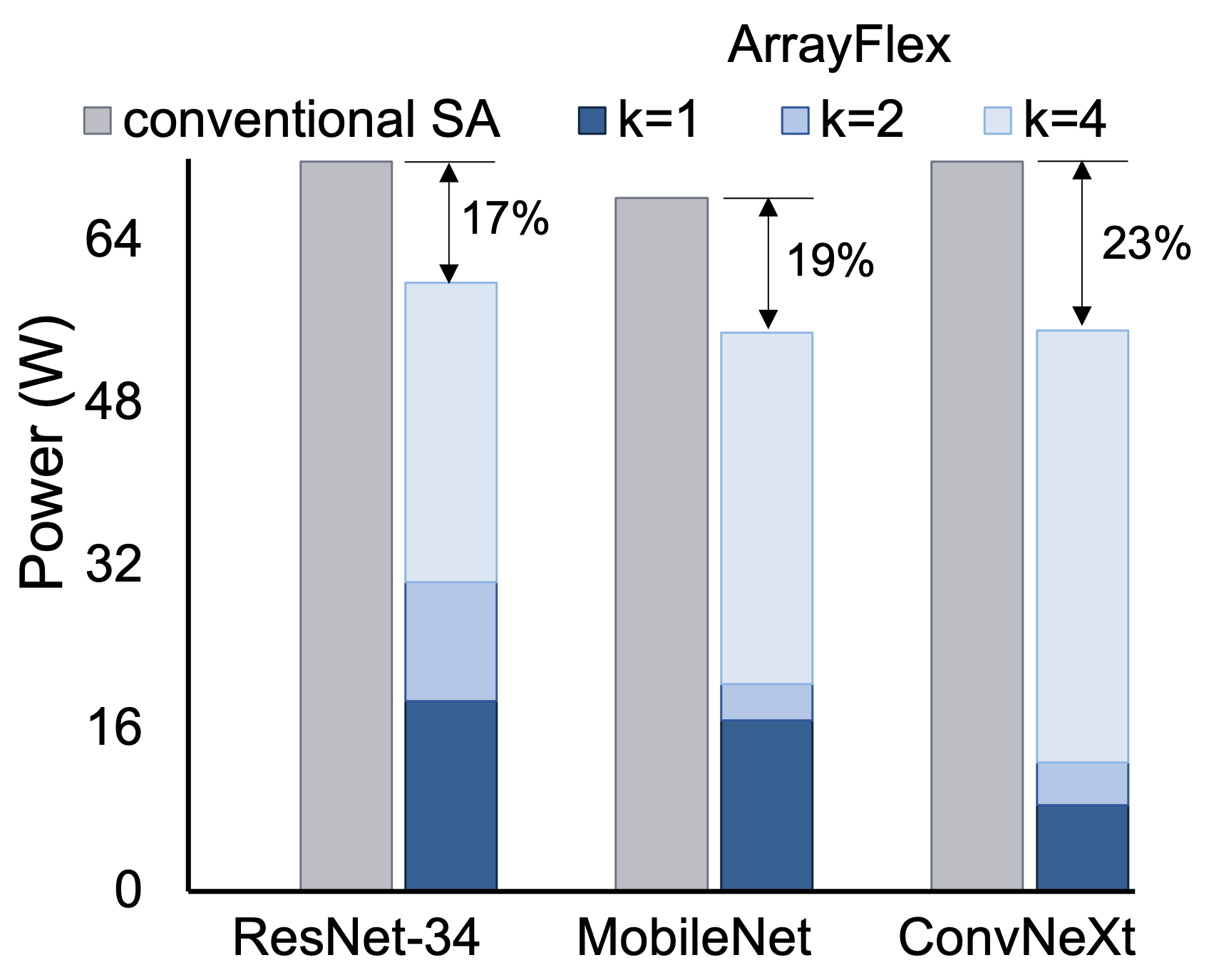} \\
{\small (a) 128$\times$128 SAs} &
{\small (b) 256$\times$256 SAs} 
\end{tabular}
\caption{The power of the SAs for \textit{complete runs} (i.e., execution of all layers) for three CNNs using (a) 128$\times$128 and (b) 256$\times$256 SAs. The power of the SRAMs and any other peripheral circuitry outside the SAs is omitted.}
\label{f:apps-power}
\end{figure}

Fig.~\ref{f:apps-power} depicts the average power consumption of both SAs under comparison when executing inference on the ResNet-34~\cite{resnet}, MobileNet~\cite{mobilenet}, and ConvNeXT~\cite{convnext} CNNs. For ArrayFlex, the power cost of each pipeline mode is shown separately. ArrayFlex operates in shallow pipeline mode in the majority of the CNN layers of each application. Consequently, this behavior translates to \textit{overall power savings} that range between 13\% and 15\% for SAs of size 128$\times$128 PEs, and increase to 17\%--23\% for SAs of size 256$\times$256 PEs. 
The combined effect of reduced power and less execution time  makes ArrayFlex 1.4$\times$--1.8$\times$ more efficient in terms of energy-delay-product than a conventional SA.

\section{Conclusions}
\label{s:conclusions}

Merging the pipeline stages of an SA creates an interesting tradeoff for the computation of GEMM. On one hand, the number of cycles required to complete the matrix multiplication is reduced proportionally to the collapsed pipeline depth. On the other hand, the clock period should increase to accommodate the larger combinational delay of the merged pipeline stages. Utilizing carry-save adders in parallel to the multiply-add components of the PEs allows us to efficiently control the clock frequency degradation. Since the clock frequency reduction is smaller than the reduction achieved in the number of cycles for certain CNN layers, the proposed ArrayFlex SA can minimize the total execution time. The reduced clock frequencies and the marginal hardware overhead incurred also allow for the reduction of power. Currently, this work focuses on dense computations. Nevertheless, since sparse layers can be mapped to GEMM blocks and executed by SAs using efficient peripheral circuitry, we plan -- as future work -- to also explore the applicability of ArrayFlex to sparse layers.

\bibliographystyle{IEEEtran}
\bibliography{refs}

\begin{thebibliography}{10}
\providecommand{\url}[1]{#1}
\csname url@samestyle\endcsname
\providecommand{\newblock}{\relax}
\providecommand{\bibinfo}[2]{#2}
\providecommand{\BIBentrySTDinterwordspacing}{\spaceskip=0pt\relax}
\providecommand{\BIBentryALTinterwordstretchfactor}{4}
\providecommand{\BIBentryALTinterwordspacing}{\spaceskip=\fontdimen2\font plus
\BIBentryALTinterwordstretchfactor\fontdimen3\font minus
  \fontdimen4\font\relax}
\providecommand{\BIBforeignlanguage}[2]{{%
\expandafter\ifx\csname l@#1\endcsname\relax
\typeout{** WARNING: IEEEtran.bst: No hyphenation pattern has been}%
\typeout{** loaded for the language `#1'. Using the pattern for}%
\typeout{** the default language instead.}%
\else
\language=\csname l@#1\endcsname
\fi
#2}}
\providecommand{\BIBdecl}{\relax}
\BIBdecl

\bibitem{convnext}
Z.~Liu \emph{et~al.}, ``{A ConvNet for the 2020s},'' in \emph{IEEE Conf. on
  Comp. Vision and Pattern Recognition (CVPR)}, 2022, pp. 11\,976--11\,986.

\bibitem{mobilenet}
A.~G. Howard \emph{et~al.}, ``Mobilenets: Efficient convolutional neural
  networks for mobile vision applications,'' \emph{arXiv:1704.04861}, 2017.

\bibitem{NLP_CNN}
T.~Young, D.~Hazarika, S.~Poria, and E.~Cambria, ``{Recent Trends in Deep
  Learning Based Natural Language Processing},'' \emph{IEEE Computational
  Intelligence Magazine}, vol.~13, no.~3, pp. 55 -- 75, 2018.

\bibitem{CNN-SLAM}
K.~Tateno, F.~Tombari, I.~Laina, and N.~Navab, ``{{CNN-SLAM}: {Real-Time Dense
  Monocular {SLAM} with Learned Depth Prediction}},'' in \emph{IEEE Conf. on
  Comp. Vision and Pattern Recognition (CVPR)}, 2017, pp. 6243--6252.

\bibitem{cudnn}
S.~Chetlur \emph{et~al.}, ``{cuDNN}: Efficient primitives for deep learning,''
  \emph{arXiv preprint arXiv:1410.0759}, 2014.

\bibitem{why-systolic}
H.~T. Kung, ``Why systolic architectures?'' \emph{Computer}, vol.~15, no.~1,
  pp. 37--46, 1982.

\bibitem{tpu}
N.~P. Jouppi \emph{et~al.}, ``In-datacenter performance analysis of a tensor
  processing unit,'' in \emph{Int. Symp. on Comp. Arch. (ISCA)}, 2017, p.
  1–12.

\bibitem{scalesim}
A.~Samajdar \emph{et~al.}, ``A systematic methodology for characterizing
  scalability of {DNN} accelerators using scale-sim,'' in \emph{IEEE Int. Symp.
  on Perf. Analysis of Systems and Software (ISPASS)}, 2020, pp. 58--68.

\bibitem{auto-sa}
X.~Wei \emph{et~al.}, ``Automated systolic array architecture synthesis for
  high throughput {CNN} inference on {FPGAs},'' in \emph{DAC}, 2017.

\bibitem{meissa}
B.~Asgari, R.~Hadidi, and H.~Kim, ``Meissa: Multiplying matrices efficiently in
  a scalable systolic architecture,'' in \emph{IEEE Int. Conf. on Computer
  Design (ICCD)}, 2020, pp. 130--137.

\bibitem{factored-sa}
I.~Ullah, K.~Inayat, J.-S. Yang, and J.~Chung, ``Factored radix-8 systolic
  array for tensor processing,'' in \emph{Design Automation Conf. (DAC)}, 2020.

\bibitem{reconfig-bitwidth}
V.~Camus \emph{et~al.}, ``Review and benchmarking of precision-scalable
  multiply-accumulate unit architectures for embedded neural-network
  processing,'' \emph{IEEE JETCAS}, vol.~9, no.~4, pp. 697--711, 2019.

\bibitem{eyriss2}
Y.-H. Chen, T.-J. Yang, J.~Emer, and V.~Sze, ``Eyeriss v2: A flexible
  accelerator for emerging deep neural networks on mobile devices,'' \emph{IEEE
  JETCAS}, vol.~9, no.~2, pp. 292--308, 2019.

\bibitem{hetero-sa}
R.~Xu \emph{et~al.}, ``Configurable multi-directional systolic array
  architecture for convolutional neural networks,'' \emph{ACM TACO}, vol.~18,
  no.~4, July 2021.

\bibitem{dataflow_mirroring}
J.~Lee \emph{et~al.}, ``Dataflow mirroring: Architectural support for highly
  efficient fine-grained spatial multitasking on systolic-array npus,'' in
  \emph{Design Automation Conf. (DAC)}, 2021, pp. 247--252.

\bibitem{planaria}
S.~Ghodrati \emph{et~al.}, ``Planaria: Dynamic architecture fission for spatial
  multi-tenant acceleration of deep neural networks,'' in \emph{IEEE Intern.
  Symp. on Microarchitecture (MICRO)}, 2020, pp. 681--697.

\bibitem{sara}
A.~Samajdar \emph{et~al.}, ``{Self Adaptive Reconfigurable Arrays: Learning
  Flexible GEMM Accelerator Configuration and Mapping-Space Using ML},'' in
  \emph{Design Automation Conf. (DAC)}, 2022, p. 583–588.

\bibitem{sparse-tpu}
X.~He \emph{et~al.}, ``Sparse-tpu: Adapting systolic arrays for sparse
  matrices,'' in \emph{ACM Intern. Conf. on Supercomputing (SC)}, 2020, pp.
  1--12.

\bibitem{edge}
R.~Hadidi, J.~Cao, M.~S. Ryoo, and H.~Kim, ``Toward collaborative inferencing
  of deep neural networks on internet-of-things devices,'' \emph{IEEE Internet
  of Things Journal}, vol.~7, no.~6, pp. 4950--4960, 2020.

\bibitem{edge-2}
S.~I. Venieris \emph{et~al.}, ``How to reach real-time ai on consumer devices?
  solutions for programmable and custom architectures,'' in \emph{IEEE ASAP},
  July 2021, pp. 93--100.

\bibitem{rnn-batch}
F.~Silfa, J.~M. Arnau, and A.~Gonzalez, ``{E-BATCH: Energy-efficient and
  high-throughput RNN batching},'' \emph{ACM TACO}, vol.~19, no.~1, 2022.

\bibitem{collapse}
H.~Shimada, H.~Ando, and T.~Shimada, ``Pipeline stage unification: a low-energy
  consumption technique for future mobile processors,'' in \emph{Int. Symp. on
  Low power electr. and design (ISLPED)}, 2003, pp. 326--329.

\bibitem{transparent}
J.~H. Choi \emph{et~al.}, ``Improved clock-gating control scheme for
  transparent pipeline,'' in \emph{ASP-DAC}, 2010, pp. 401--406.

\bibitem{risset}
T.~Risset, ``A method to synthesize modular systolic arrays with local
  broadcast facility,'' in \emph{IEEE ASAP}, 1992, pp. 415--428.

\bibitem{resnet}
K.~He \emph{et~al.}, ``Deep residual learning for image recognition,'' in
  \emph{IEEE CVPR}, 2016, pp. 770--778.

\end{thebibliography}

\end{document}